\documentclass[twoside]{article}
\usepackage{fleqn,espcrc2,epsf}

\usepackage{graphicx}
\usepackage[figuresright]{rotating}

\newcommand{\AmS}{{\protect\the\textfont2
  A\kern-.1667em\lower.5ex\hbox{M}\kern-.125emS}}

\title{Spin splittings among charmed hadrons}

\author{Randy Lewis\address{Department of Physics, University of Regina,
        Regina, SK, S4S 0A2, Canada},
        Nilmani Mathur\address{TRIUMF, 4004 Wesbrook Mall, Vancouver, BC,
        V6T 2A3, Canada},
    and R. M. Woloshyn$^{\rm b}$}
       
\begin{document}

\begin{abstract}
The mass differences between spin-1/2 and spin-3/2 baryons are compared 
to the mass differences between spin-0 and spin-1 mesons.
Results of simulations for charmed hadrons in the quenched approximation
from a tadpole-improved anisotropic action are discussed in the
context of other lattice calculations,
quark model predictions, heavy quark symmetry
predictions and experimental data.
\vspace{-2mm}
\end{abstract}

\maketitle

\section{MOTIVATION}

The mass differences between the lightest vector and pseudoscalar mesons
containing one or two heavy quarks are persistently smaller in quenched lattice
simulations than the experimental values\cite{chbary,hypmb,hypmeson}.
Studies of
the unquenched theory have found that the discrepancy persists\cite{unquench}.

Fewer lattice simulations have been performed for the mass differences between
spin-3/2 and spin-1/2 baryons containing one or more heavy
quarks\cite{chbary,hypmb,hypbaryon,othertalk}.
It is useful to compare these baryon mass differences
to the meson mass differences mentioned above, since both are colour hyperfine
(spin-spin) effects.  Quantitative relationships between the meson and baryon
splittings have been claimed using experimental data, quark models or heavy
quark effective theory.  If these relations are respected by lattice QCD
simulations then the baryon spin splittings will be smaller than experiment
(just like mesons), but if lattice results for the baryons agree with
experiment then the relations must clearly be violated.  In either case,
the information will be helpful when seeking to understand the existing
discrepancy between lattice QCD simulations and experiment.

\section{PHENOMENOLOGY}

A simple quark model for the hadron masses uses the operators
\begin{eqnarray}
M_{\rm meson} &=& m_1 + m_2 + a\frac{\vec\sigma_1\cdot\vec\sigma_2}{m_1m_2}, \\
M_{\rm baryon} &=& m_1 + m_2 + m_3 + \sum_{i>j}\frac{a_{ij}^\prime}{2}
                               \frac{\vec\sigma_i\cdot\vec\sigma_j}{m_im_j}.
\end{eqnarray}
It follows that the spin splittings for heavy-light mesons and for singly
and doubly heavy baryons vanish like $1/m_Q$,
\begin{equation}
M_{3/2}-M_{1/2}=\frac{3c^\prime}{4c}(M_v-M_p) \sim \frac{1}{m_Q},
\end{equation}
where $c$ and $c^\prime$ contain wave function information.  The determination
of $c^\prime/c$, using a relation between the wave function at the origin
and the derivative of the potential (plus some other postulates), has been
discussed by Lipkin\cite{Lipkin} and by Lipkin and O'Donnell\cite{LipO}.

\begin{table}[thb]
\caption{Hadron creation operators.}\label{ops}
\begin{tabular}{r@{ = }l}
   \hline
   \multicolumn{2}{c}{mesons} \\
   $D$ & $Q^\dagger\gamma_5u$ \\
   $D^*$ & $Q^\dagger\gamma_\mu u$ \\
   \hline
   \multicolumn{2}{c}{spin-1/2 baryons} \\
   $\Lambda_Q$ & $\frac{1}{\sqrt{6}}\epsilon^{abc}\left(2[u_a^TC\gamma_5d_b]Q_c
              + [u_a^TC\gamma_5Q_b]d_c\right.$ \\
   \multicolumn{2}{c}{$\left.-[d_a^TC\gamma_5Q_b]u_c\right)$} \\
   $\Sigma_Q$ & $\epsilon^{abc}[u_a^TC\gamma_5Q_b]d_c$ \\
   $\Xi_Q$ & $\epsilon^{abc}[u_a^TC\gamma_5Q_b]s_c$ \\
   $\Xi_Q^\prime$ & $\frac{1}{\sqrt{2}}\epsilon^{abc}
              \left([u_a^TC\gamma_5Q_b]s_c+[s_a^TC\gamma_5Q_b]u_c\right)$ \\
   $\Omega_Q$ & $\epsilon^{abc}[s_a^TC\gamma_5Q_b]s_c$ \\
   $\Xi_{QQ}$ & $\epsilon^{abc}[u_a^TC\gamma_5Q_b]Q_c$ \\
   $\Omega_{QQ}$ & $\epsilon^{abc}[s_a^TC\gamma_5Q_b]Q_c$ \\
   \hline
   \multicolumn{2}{c}{spin-3/2 baryons} \\
   $\Sigma_Q^*$ & $\epsilon^{abc}[u_a^TC\gamma_\mu d_b]Q_c$ \\
   $\Xi_Q^*$ & $\epsilon^{abc}[u_a^TC\gamma_\mu s_b]Q_c$ \\
   $\Omega_Q^*$ & $\epsilon^{abc}[s_a^TC\gamma_\mu s_b]Q_c$ \\
   $\Xi_{QQ}^*$ & $\epsilon^{abc}[Q_a^TC\gamma_\mu Q_b]u_c$ \\
   $\Omega_{QQ}^*$ & $\epsilon^{abc}[Q_a^TC\gamma_\mu Q_b]s_c$ \\
   \hline
\end{tabular}
\vspace{-5mm}
\end{table}

For doubly heavy baryons the pair of heavy quarks forms a colour
anti-triplet, so $c\approx c^\prime$ and
\begin{eqnarray}\label{QQmodel}
\left[M_{3/2}-M_{1/2}\right]_{QQq} = \frac{3}{4}(M_v-M_p).
\end{eqnarray}
A more elegant derivation of this same result comes from incorporating the
anti-triplet diquark directly into heavy quark effective
theory\cite{SavWis}.
The resulting Lagrangian has a superflavour symmetry among heavy quarks and
heavy diquarks:
\begin{eqnarray}
{\cal L} \!\!\! &=& \!\!\!
         \chi^\dagger\left[iD^0+\frac{\lambda\vec{D}^2}{2m_Q}+\frac{g_s
         \vec\Sigma\cdot\vec{B}}{2m_Q}+O\left(\frac{1}{m_Q^2}\right)\right]\chi
         \nonumber \\
\end{eqnarray}
where
\begin{eqnarray}
\chi^T &=& (Q_\uparrow,Q_\downarrow,\{QQ\}_{+1},\{QQ\}_{0},\{QQ\}_{-1}), \\
\lambda &=& {\rm diag}\left(1,1,\frac{1}{2},\frac{1}{2},\frac{1}{2}\right),
\end{eqnarray}
and $\vec\Sigma$ is the block-diagonal spin vector for quarks and diquarks.
A direct calculation from this Lagrangian leads to Eq.~(\ref{QQmodel}).

\section{LATTICE CHOICES AND METHOD}

As reported in Ref.~\cite{chbary}, our simulations used an anisotropic
tadpole-improved relativistic action for gauge fields and for quarks.
The gauge action includes $1\times2$ rectangular plaquettes which remove
$O(a^2)$ classical errors, and the D234 quark action has no $O(a)$ nor
$O(a^2)$ classical errors.

The hadron creation operators are listed in Table~\ref{ops}.  Sink-smeared
and local correlators were built from these operators and
fit simultaneously to single and double exponential functions.
The ``spin-3/2'' operators in Table~\ref{ops} actually contain an admixture
of spin-1/2 as well.  The spin-3/2 component is isolated by
a judicious choice of Lorentz components.

\begin{table}[thb]
\caption{Lattice parameters.}\label{details}
\begin{tabular}{cccc}
$\beta$ & size & \# of config's & $a_t^{-1}({\rm GeV})$ \\
\hline
2.1 & $12^3\times32$ & 720 & 1.803(42) \\
2.3 & $14^3\times38$ & 442 & 2.210(72) \\
2.5 & $18^3\times46$ & 325 & 2.625(67) \\
\hline
\end{tabular}
\vspace{-7mm}
\end{table}

In our simulations, the bare anisotropy was set to $a_s/a_t=2$, tadpole
improvement was implemented via the mean link in Landau gauge, and
quarks have Dirichlet time boundaries.
Configuration details are given in Table~\ref{details}.

\section{COMPUTED HADRON MASSES}

\begin{figure}[thb]
\vspace{-5mm}
\epsfxsize=220pt \epsfbox[25 300 612 685]{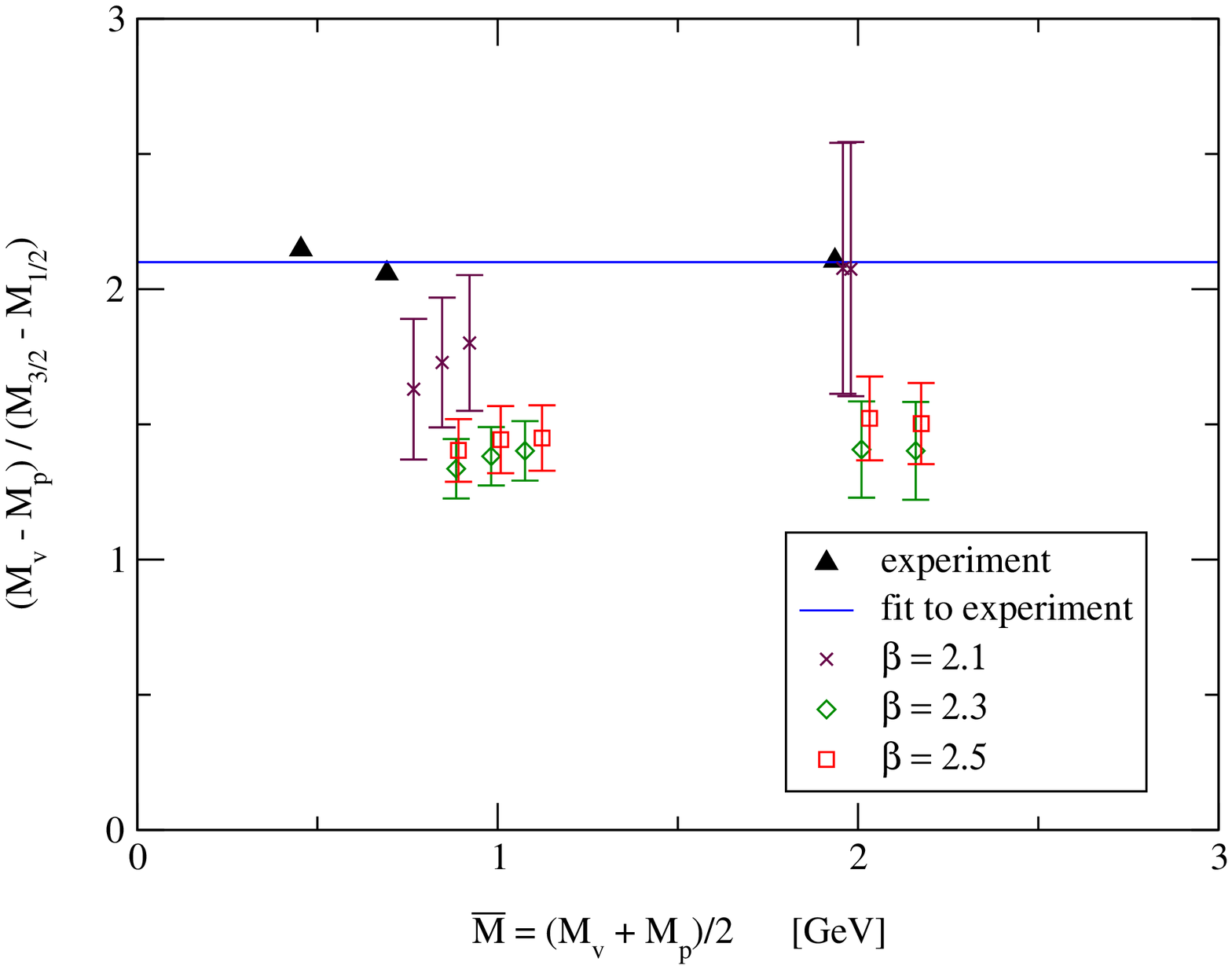}
\caption{Ratio of spin splittings: mesons/singly heavy baryons.}
\label{fig:ratiosingly}
\vspace{5mm}
\epsfxsize=220pt \epsfbox[25 300 612 685]{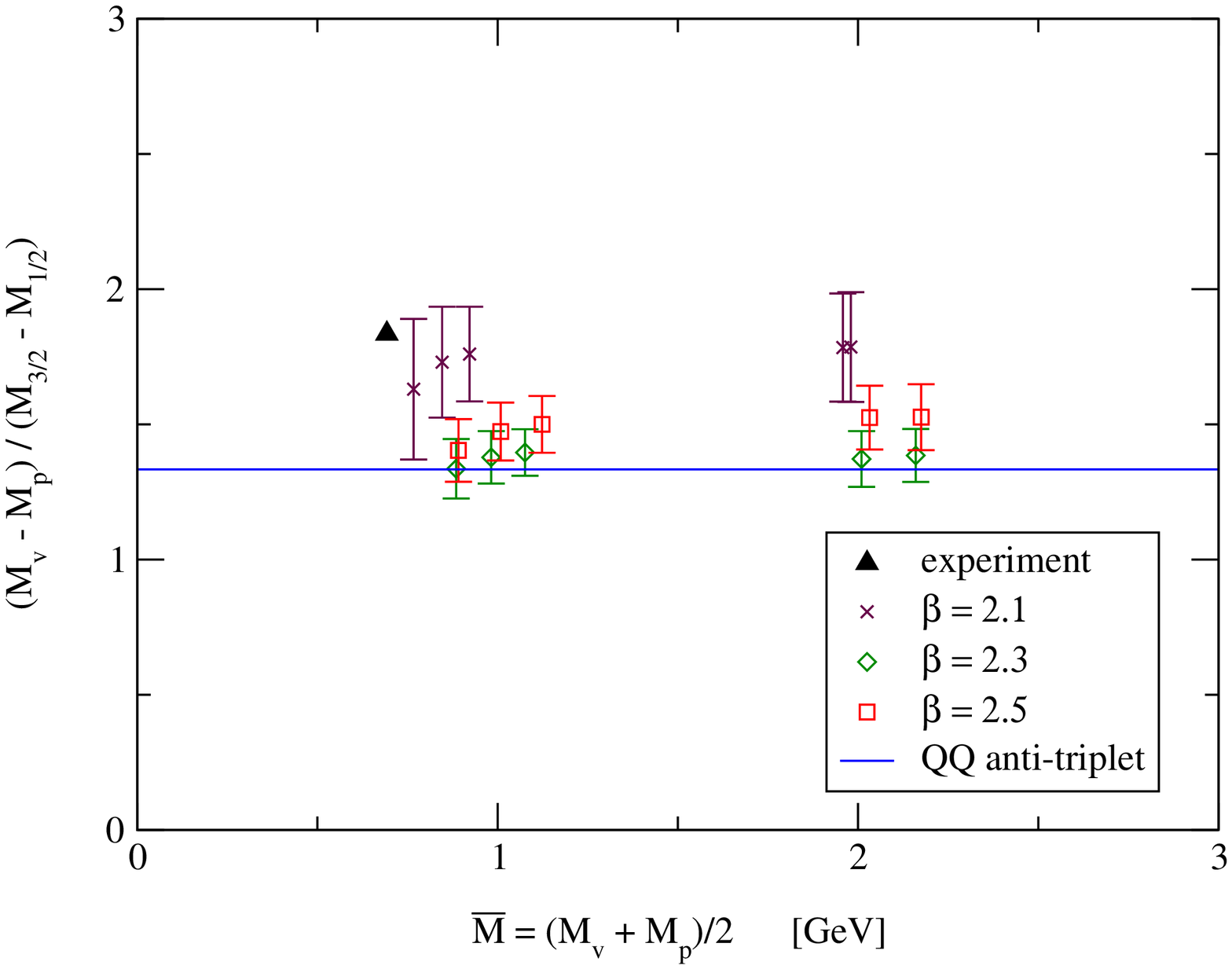}
\caption{Ratio of spin splittings: mesons/doubly heavy baryons.}
\label{fig:ratiodoubly}
\vspace{-5mm}
\end{figure}

As shown in Fig.~\ref{fig:ratiosingly}, the experimental ratio of
$\bar{Q}q$ meson spin splittings to $Qqq$ baryon spin splittings is
remarkably constant, where $q$ is a light quark and $Q$ is light or strange
or charmed.  Our lattice results, without any extrapolation in $a$ or $\kappa$,
are also independent of $m_Q$, but significantly smaller than experiment.

Fig.~\ref{fig:ratiodoubly} shows the ratio of spin splittings for $\bar{Q}q$
mesons and $QQq$ baryons.  Once again, the lattice results are independent of
$m_Q$ and smaller than the experimental point.  Interestingly, they are near
the prediction of Eq.~(\ref{QQmodel}).

\begin{figure}[thb]
\vspace{-5mm}
\epsfxsize=220pt \epsfbox[25 300 612 685]{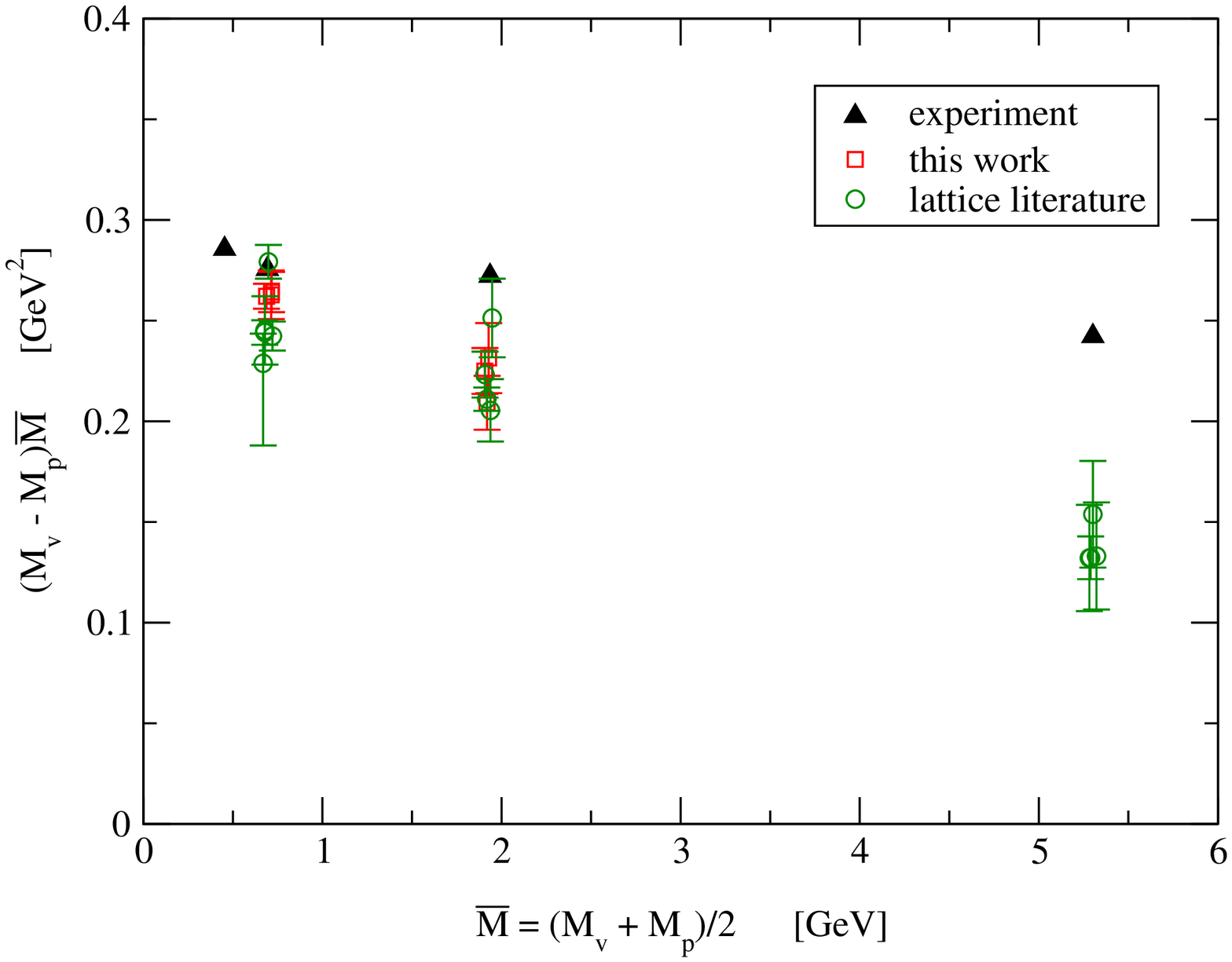}
\caption{Spin splittings among heavy-light mesons.  ``Lattice literature'' is
         Refs.~\protect\cite{hypmb,hypmeson}.}
\label{fig:mesons}
\vspace{5mm}
\epsfxsize=220pt \epsfbox[25 300 612 685]{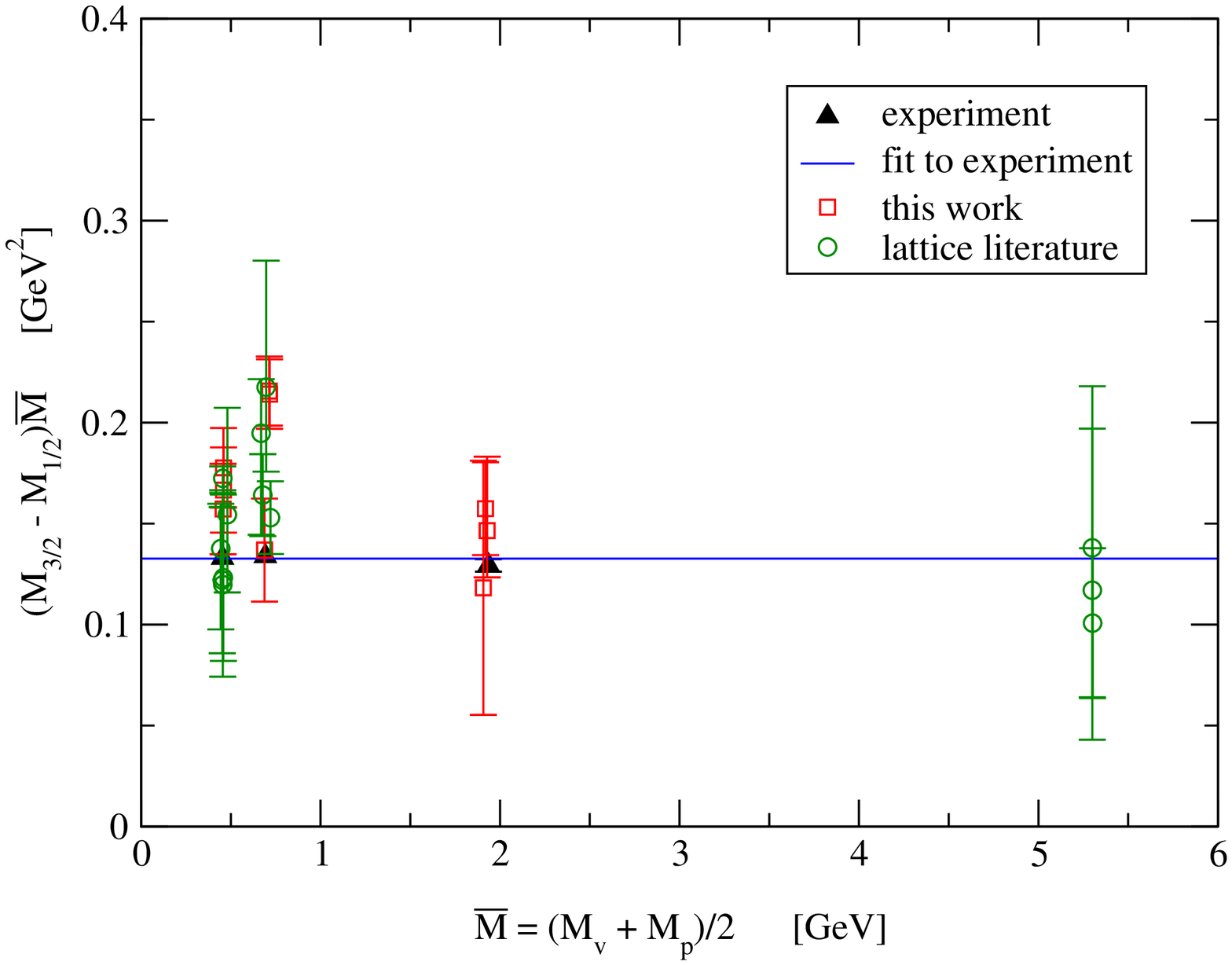}
\caption{Spin splittings among singly heavy baryons.  ``Lattice literature'' is
         Refs.~\protect\cite{hypmb,hypbaryon,othertalk}.}
\label{fig:singly}
\vspace{-5mm}
\end{figure}

Some representative lattice computations of the heavy-light meson spin
splittings are shown in Fig.~\ref{fig:mesons}, for $Q$ from light quarks
through to
bottom.  The experimental results are linear in the average mesons mass to
a few percent, but the lattice results show a significant slope.  The lattice
bottom splitting is about half of the experimental value.
Heavy quark symmetry requires that $(M_v-M_p)\bar{M} \to $ constant as
$\bar{M} \to \infty$, but Fig.~\ref{fig:mesons} does not allow us to determine
what the heavy quark limiting value is.

In contrast to the meson situation, the lattice results for the $Qqq$ baryon
spin splittings,
plotted in Fig.~\ref{fig:singly}, are not significantly smaller
than experiment; in fact, some lattice results are larger than experiment.

\section*{ACKNOWLEDGEMENTS}

This work was supported in part by the Natural Sciences
and Engineering Research Council of Canada.
Some of the computing was done on hardware funded by the Canada Foundation for
Innovation, with contributions from Compaq Canada, Avnet Enterprise Solutions,
and the Government of Saskatchewan.

\end{document}